\documentclass{article}
\usepackage{spconf,amsmath,epsfig}

\providecommand{\doi}[1]{doi: {\footnotesize \href{http://dx.doi.org/#1}{\path{#1}}}}

\usepackage[pdftex=true,breaklinks=true,hidelinks=true,colorlinks=true,citecolor=blue]{hyperref}

\usepackage[square,numbers]{natbib}
\usepackage{textalpha}
\usepackage{natbib}

\usepackage{multicol}

\usepackage{graphicx}

\title{AI4FAPAR: how artificial intelligence can help to forecast the seasonal earth observation signal}
\name{Filip Sabo, Martin Claverie, Michele Meroni, Arthur Hrast Essenfelder.}

\address{European Commission, Joint Research Centre (JRC), Ispra, Italy.}
\begin{document}
%
\maketitle
\begin{abstract}
This paper investigated the potential of a multivariate Transformer model to forecast the temporal trajectory of the Fraction of Absorbed Photosynthetically Active Radiation (FAPAR) for short (1 month) and long horizon (more than 1 month) periods at the regional level in Europe and North Africa. The input data covers the period from 2002 to 2022 and includes remote sensing and weather data for modelling FAPAR predictions. The model was evaluated using a leave one year out cross-validation and compared with the climatological benchmark. Results show that the transformer model outperforms the benchmark model for one month forecasting horizon, after which the climatological benchmark is better. The RMSE values of the transformer model ranged from 0.02 to 0.04 FAPAR units for the first 2 months of predictions. Overall, the tested Transformer model is a valid method for FAPAR forecasting, especially when combined with weather data and used for short-term predictions.
\end{abstract}
\begin{keywords}
Satellite Remote Sensing, Agricultural Monitoring, Deep Learning, Forecasting, FAPAR
\end{keywords}
\section{Introduction}
\label{sec:intro}
Earth observation (EO) provides invaluable information for monitoring crop conditions. Leveraging EO data can enhance the effectiveness of crop monitoring by assessing its productivity and quality \cite{defourny2019near}. This contributes to ensuring a stable food supply and minimizing the adverse effects of agricultural practices. The availability of continuous and long-term EO data, captured at a moderate spatial resolution, offers substantial benefits for modeling crop yields, as it offers a long term data record capturing the multi-annual variability of crop conditions and yields, suitable for training data-driven approaches.  

Currently available operational systems for crop monitoring, such as Anomaly Hotspots of Agricultural Production (ASAP) or Monitoring Agricultural ResourceS (MARS) Bulletin \cite{VANDERVELDE201956}, heavily rely on EO and agrometeorological data to deliver pertinent information to crop analysts. Indicators of biomass status such as Fraction of Absorbed Photosynthetically Active Radiation (FAPAR) are considered as proxies of crop yield and therefore the temporal evolution of FAPAR is monitored during the growing season. Specifically, FAPAR measures the proportion of photosynthetically active radiation in the 0.4–0.7 μm spectrum that is absorbed by leaves \cite{FENSHOLT2004490}.
Forecasting FAPAR for the near future could enhance the timeliness and utility of the early warning information. Deep learning (DL) models can provide an improved method for anticipating the detection of vegetation anomalies by learning the lagged response of FAPAR to weather data based on long-term time series EO and weather data. Recently, several studies have tested different DL models for forecasting vegetation dynamics over different areas \cite[e.g.,][]{lees_deep_2022, cui_forecasting_2020, ahmad_machine-learning_nodate}. 

Authors in \cite{lees_deep_2022} used long short term memory (LSTM) models for short term forecasting (one month) of vegetation health in Kenya. Convolutional neural network (CNN) was proposed by \cite{cui_forecasting_2020} for forecasting the normalized difference vegetation index (NDVI) for the next three months with a 30 day interval (3 steps ahead prediction) in multiple regions in North America and Colombia. In order to benefit from spatial and temporal features, authors in \cite{ahmad_machine-learning_nodate} combined the CNNs with LSTM for predicting the next NDVI value over soybean pixels. 

Despite the successful applications of DL models, there is still a lack of studies that use sequence-to-sequence (S2S) models for EO time series (TS) forecasting. We aim to bridge this literature gap by introducing and evaluating a Transformer model for short (one month) and long-term (seasonal) TS forecasting at the regional level in Europe and North Africa using remote sensing and weather data. This study will contribute to the development of more accurate and reliable early warning systems for monitoring vegetation health and productivity. 

\label{sec:expe}
\begin{figure*}[htb]
  \centering
  \includegraphics[width=.82\paperwidth]{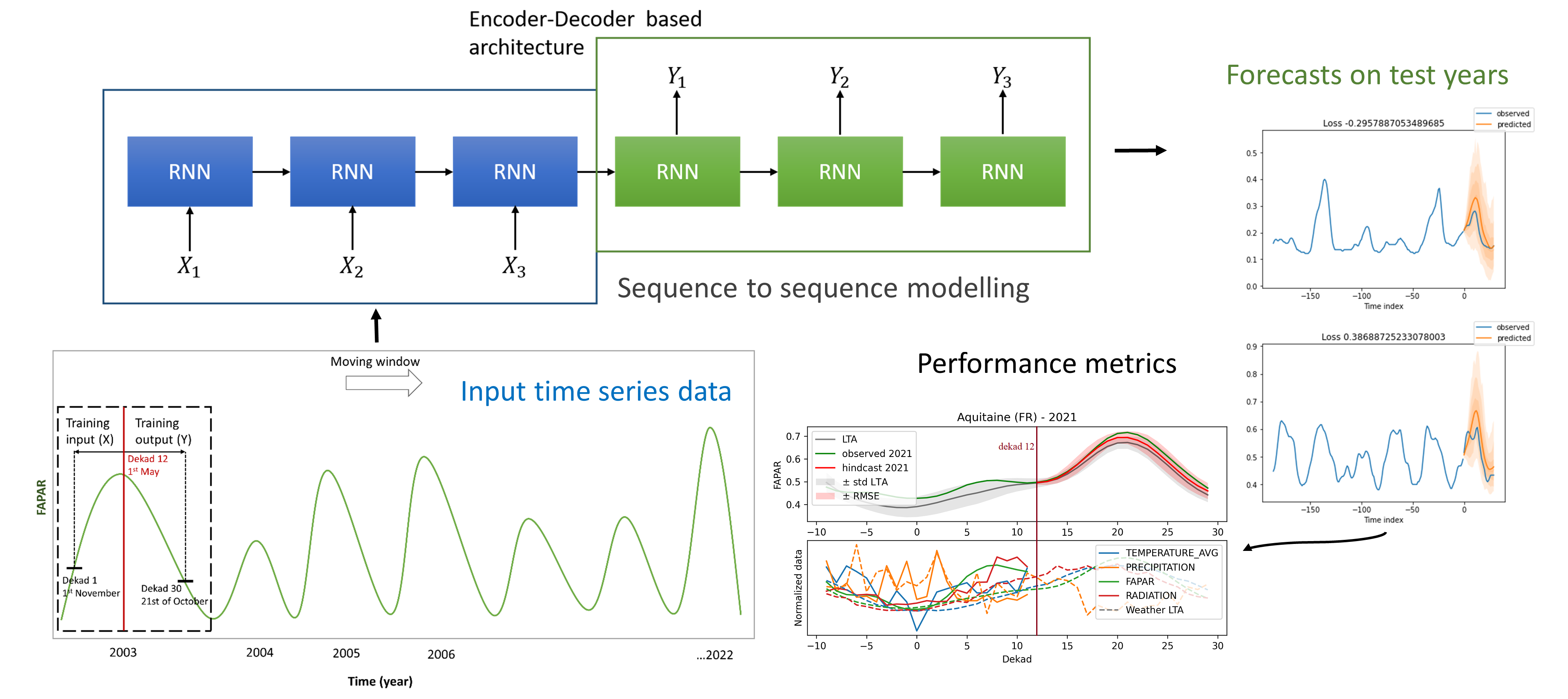}
  \caption{A simplified processing workflow. Input time series data are first prepared for training with a moving window approach (X and Y). The batched data are then passed to the Temporal Fusion Transformer for S2S modelling. The modelling is then followed by the prediction on unseen test year in a leave-one-year out approach. 
  Weather data preparation is not shown for clarity. This workflow repeats for each training step increment.}
  \label{fig:encoder}
\end{figure*}
\section{Sequence to Sequence models}
\label{sec:method}

Recent deep learning models for time series forecasting can be classified into univariate and multivariate approaches. Univariate models are usually autoregressive models which means that they model the future values based on their past values  \cite{challu_n-hits_2022, salinas_deepar_2020}. Optionally, the univariate models can employ covariates which are known both in the future and in the past, for example the time variable. Contrary to the univariate models, the multivariate models such is the Temporal Fusion Transformer (TFT) \cite{lim_temporal_2021} can employ exogenous continuous (weather and time data) and categorical static variables (region ID) to better model the future predictions based on meteorological drivers of vegetation growth up to the time of forecast. TFT is a combination of transformers and LSTM modules that can handle both sequential and non-sequential data and it has been designed to learn from multiple related time series to make accurate predictions for a target variable. These univariate and multivariate variants are also called sequence-to-sequence (S2S) models with multiple horizon forecasts. The architectures of S2S models, encoder-decoder based, allow variable input and output size, thus allowing both short- and long-term horizon forecasts and can be used to predict an entire forecast horizon at once.

\begin{table}[]
\caption{Input data. LTA corresponds to long term average.}
\label{tab:input}
\centering
\begin{tabular}{lll}
\hline
Past covariates     & Future covariates       &  \\
\hline
FAPAR               & FAPAR LTA               &  \\
Average temperature & Average temperature LTA &  \\
Precipitation sum   & Precipitation sum LTA   &  \\
Global radiation    & Global radiation LTA    &  \\
Growing degree days & Growing degree days LTA &  \\
Time index (dekad)  & Time index (dekad)      &  \\
Region ID           & Region ID               &  \\
\hline
\end{tabular}
\end{table}

For the purpose of long- and short-horizon forecasting of FAPAR values, the TFT model was selected because of the following properties: (1) it is a S2S model; (2) it is multivariate and it allows static categorical inputs; (3) it accepts past and future known covariates; and (4) it is interpretable due to its attention mechanisms.
We benchmark the performance of the TFT model against a null model which uses long term average (LTA) values as predictions.

\section{Data}
\label{sec:data}

The input used in this study relies on the MARS crop yield forecasting system (MCYFS) data set, covering 33 countries: all 27 EU member states, UK, Ukraine, Morocco, Algeria and Tunisia. 

The FAPAR data were extracted from the remote sensing component (RS) of the MCYFS. The product relies on the MODIS FAPAR product MCD15A2H version 6.1 at 500m spatial resolution, with a 4-days time step. The FAPAR workflow consists in downloading the data, compositing in time the product at a dekadal time step (every 10 days) and applying a smoothing algorithm. The latter relies on Whittaker algorithms and allows to gap-fill all missing data due mostly to cloud cover. A special attention is provided on near real-time data where extrapolation of the signal is constrained  by using statistics of the trend derived from the pixel-based long-term record \cite{MERONI2019508}. 

The gap-filled pixel-based dekadal maps are then aggregated at NUTS-2 administrative level  using arable land mask derived from the Copernicus CORINE Land Cover 2018 (https://land.copernicus.eu/pan-european/corine-land-cover/clc2018). The NUTS-2 level, roughly corresponding the regional level, is commonly used by analysts to summarise the information (e.g. in the MARS Bulletin). For the 33 countries, this results in 359 administrative regions.

Over the same administrative regions, weather information is extracted and aggregated to the same spatial and temporal frame as the RS data. To predict FAPAR we use the following five variables as predictors: average temperature, growing degree days, precipitation sum, solar radiation, time index (dekad number) and region id. 
Table \ref{tab:input} summarizes the input past and future covariates used in the study. 

\section{EXPERIMENTAL SETUP}

The multivariate time series data (FAPAR, average temperature, solar radiation, precipitation and growing degree days) spans from 2002 until 2022. The data was split into sequences of shorter length (i.e. batches). The two windows X,Y represent the desired look-back (encoder length) and the forecast horizon (decoder length), respectively (training input X and output Y in fig. \ref{fig:encoder}). That is, 18 dekads (6 months) of multivariate time series data was passed to the model as features and 18 dekads of FAPAR values were used as the label (target variable) and this sequence repeats for the entire time range (2003-2022 cropping seasons). The simplified workflow is shown in fig. \ref{fig:encoder}. Time series data split is shown only for FAPAR due to clarity.

Errors statistics were evaluated using leave one year out cross-validation. That is, one year at a time was removed from the data set and used to evaluate the prediction accuracy of the model. At each loop of the cross-validation, the remaining n-1 years were used to train the model.

TFT can be fed with past covariates (i.e. time series available only up to the time of forecast and thus unknown for the future) and future covariates (i.e. time series with values known for the future as well). The future covariates can include data such as month, day of the year, dekad or weather forecasts. TFT will take future covariates into account up to the forecast horizon when making forecasts. In this study LTA values of all the TS variables were used also as future known covariates. Dekad number was passed as a time variable and region id was passed as a static categorical variable. The past covariates are fed to the encoder part of the TFT while the future covariates are passed to both the encoder and decoder modules of the network.
\label{sec:results}
\begin{figure}[ht]
  \centering
  \includegraphics[width=\columnwidth]{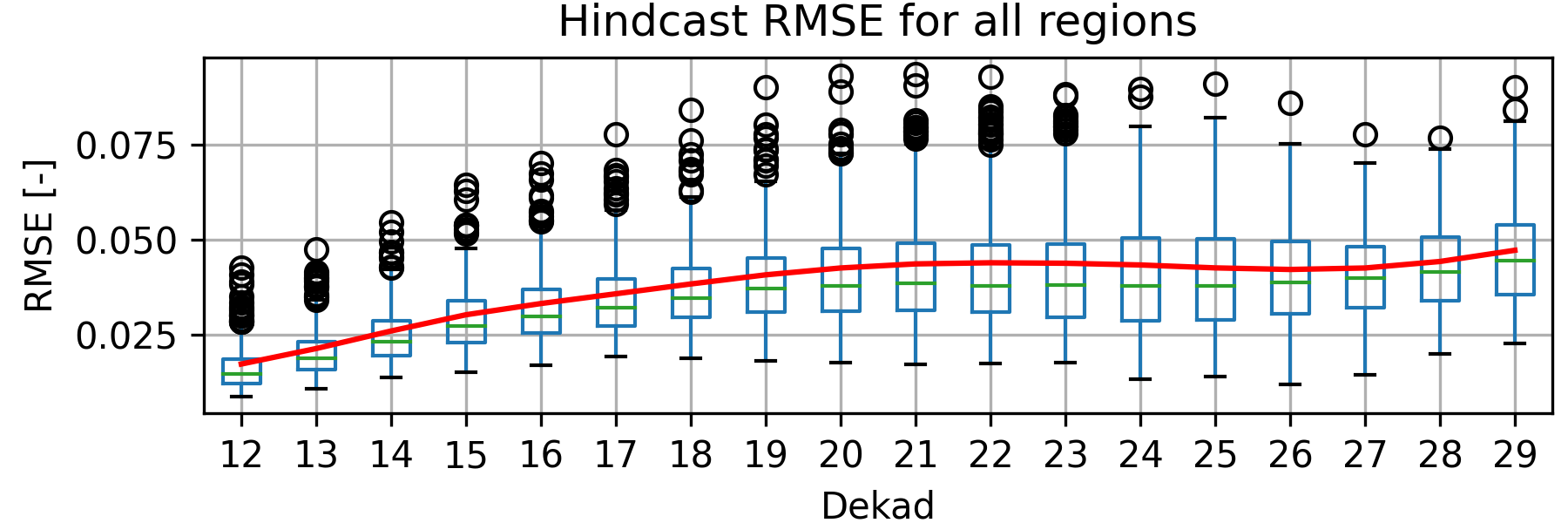}
  \caption{Boxplots of hindcasted FAPAR RMSE for all regions. RMSE is computed per region using the 2003-2022 hindcasts. Red line represents the RMSE computed with all regions together.}
  \label{fig:boxplotRMSE}
\end{figure}

\section{Results}

The hindcasted results were evaluated per dekad and per region by calculating the Root Mean Square Error (RMSE). The predicted sequence contained 18 values, that is, 18 dekads or 6 months, for each region and for each cropping season (2003-2022). Predictions start from dekad 12 (1st of May) and end on the 21st of October (dekad 29). Fig. \ref{fig:boxplotRMSE} shows the RMSE boxplots and the RMSE values calculated for all regions together. The RMSE values ranged from 0.02 to 0.04 for the predictions of the first two months and then increased a little over 0.04 for the rest of the predicted sequence. The RMSE values remain stable and do not fluctuate much after the second month (after dekad 19).

\begin{figure}[h]
  \centering
  \includegraphics[width=\columnwidth]{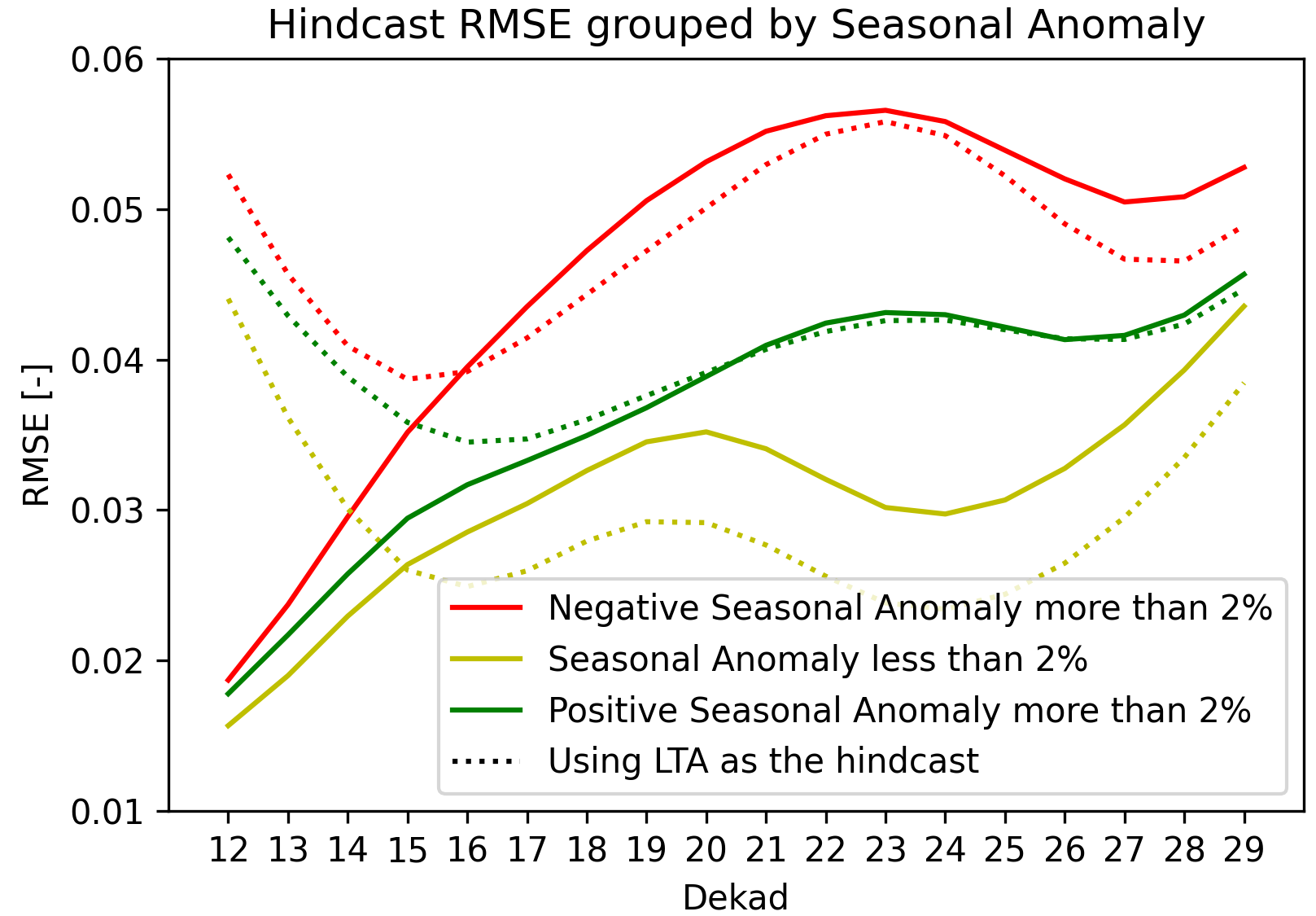}
  \caption{Hindcasted FAPAR RMSE grouped in three category according the seasonal anomaly of the hindcast period, i.e. the relative difference between observed FAPAR and LTA. The three categories encompass between 30 and 37\% of all the Region/Season pairs. Dashed lines corresponds to RMSE if the hindcast is replaced by the LTA.}
  \label{fig:RMSE_Seasonal_Anomaly}
\end{figure}

The RMSE of the TFT model is compared with that of the LTA benchmark in fig. \ref{fig:RMSE_Seasonal_Anomaly} for the 3 classified seasonal anomalies: $\le-2\%$, [-2\%,2\%],$\ge2\%$, corresponding to the relative differences between the observed FAPAR and the LTA during the forecast period (dekads 12-29). RMSE of the TFT classified anomalies are shown with a solid line while the dashed lines represent the RMSE of the LTA anomalies. The TFT model demonstrates better performance than the LTA in the first 4 dekads (12-15) for the negative seasonal anomaly class. For the positive anomaly class, the RMSE vales of the TFT are lower for the first 7 dekads. In other words, the model outperformed the LTA for predictions of more than 2 months. In the seasonal anomaly of less than 2 percent group, the TFT model was better than the benchmark for the first 3 dekads (1 month). For the the rest of the predicted sequence (more than 2 months) the LTA benchmark was better than the TFT model.

\begin{figure}
  \centering
  \includegraphics[width=\columnwidth]{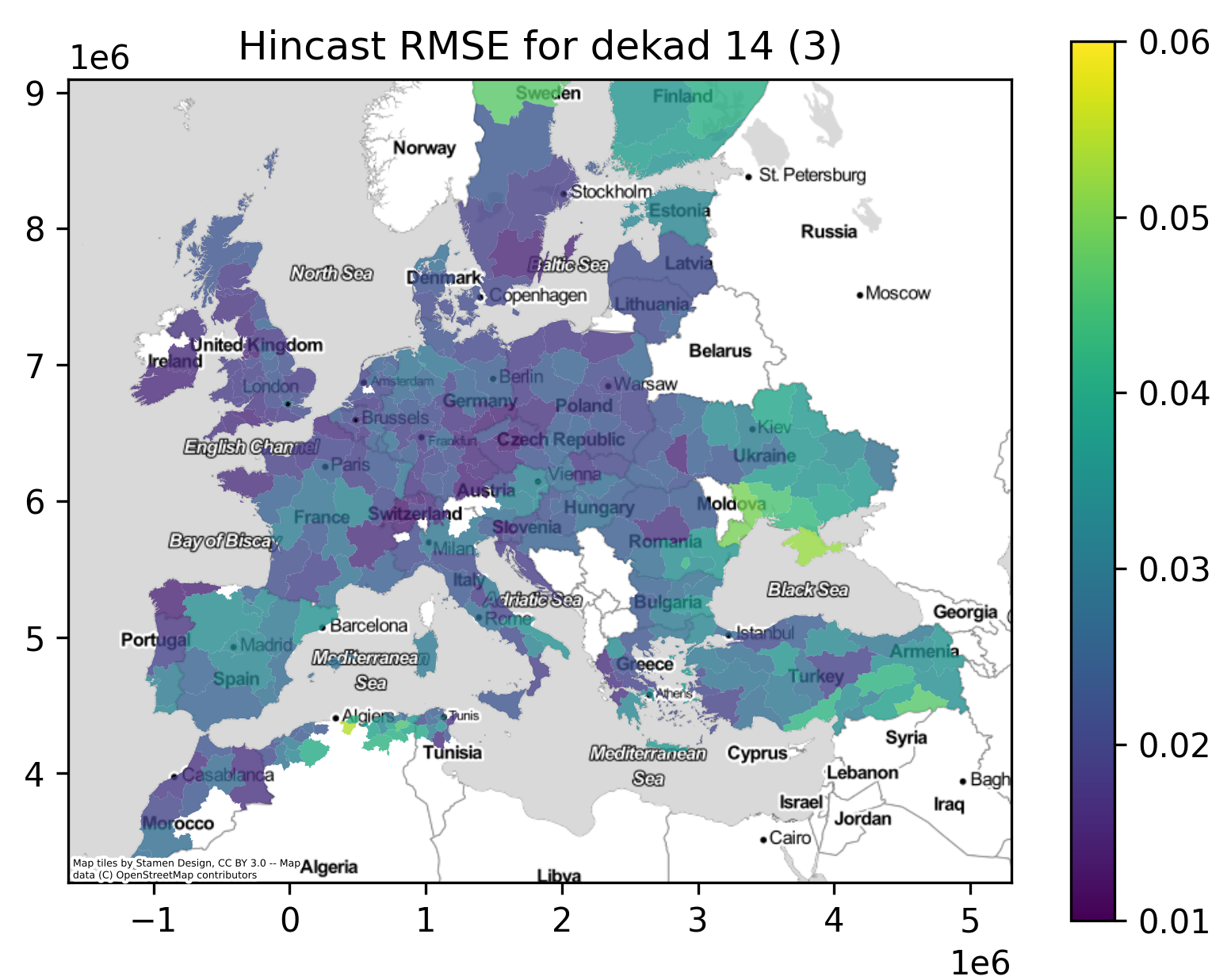}
  \caption{Map of the Hindcast RMSE for dekad 14 (3 dekads after last Observation).}
  \label{fig:MapRMSE}
\end{figure}

In fig. \ref{fig:MapRMSE} it can be observed that for the RMSE of the 3rd forecasted there is little spatial variation.

\section{CONCLUSIONS AND DISCUSSIONS}
\label{sec:conclusion}

This study introduced a workflow for short- and long-horizon forecasting of FAPAR using a multivariate transformer model and weather data at regional level in Europe and North Africa. The model evaluation showed that the RMSE of the predictions ranged from 0.02 to 0.04 for all the regions. The RMSE for the two month prediction horizon was mainly below 0.04. An added value of the model compared to the LTA was observed for the anomaly classification for the short-term forecasts, up to one month. For the long-term forecasts, the LTA was more advantageous. Future experimental directions will be to use pixel-based sampling instead of regional aggregations of the input TS data and to explore other models based on auto-regressive CNN.

\bibliographystyle{plainnat}
\bibliography{ndvi_forecast.bib}

\small



\end{document}